\documentclass[preprint]{ptephy_v1}%%%%%% to generate preprint number with ptep logo

\preprintnumber{XXXX-XXXX} %%% %%% Insert preprint number here
\usepackage{hyperref}

\newcommand{\be}{\begin{equation}}
\newcommand{\ee}{\end{equation}}
\newcommand{\bea}{\begin{eqnarray}}
\newcommand{\eea}{\end{eqnarray}}
\newcommand{\ba}{\begin{array}}
\newcommand{\ea}{\end{array}}

\newcommand{\cL}{\mathcal{L}}

\newcommand{\cB}{\mathcal{B}}
\newcommand{\cM}{\mathcal{M}}
\newcommand{\UU}{{\rm U}}
\newcommand{\SU}{{\rm SU}}
\def\heavy{{[3]}}
\def\light{{[12]}}

\begin{document}

\title{Flavor physics beyond the Standard Model \\
and the Kobayashi-Maskawa legacy}

%%%% To generate auto affiliation numbers please use \author{}\affil{} command

\author{Gino Isidori}
\affil{Physik-Institut, Universit\"at Z\"urich, CH-8057, Switzerland \email{gino.isidori@uzh.ch}}

%%% To include the collaborator name... Please use the command "\collaborator"
%%% For example: \collaborator{ATLAS Collaboration}

\begin{abstract}
The Kobayashi-Maskawa (KM) hypothesis about the existence of a third generation of quarks represents a cornerstone of the Standard Model (SM). Fifty years after this seminal paper, flavor physics continues to represent a privileged observatory on physics occurring at high energy scales.
In this paper I first review this statement using general effective-theory arguments, highlighting some interesting modern lessons from the KM paper. I then discuss some novel extensions of the SM based on the concept of flavor deconstruction: the hypothesis that gauge interactions are manifestly flavor non universal in the ultraviolet. The phenomenological consequences of this class of models are also briefly illustrated.
\end{abstract}

\maketitle

\section{Introduction}
In 1973, before the experimental discovery of charm,  Kobayashi and Maskawa (KM)~\cite{Kobayashi:1973fv} proposed the existence of a third generation of chiral fermions to explain the phenomenon of CP violation observed in the kaon system. This hypothesis, which was definitely bold at the time, turned out to be a cornerstone of our present understanding of fundamental interactions.  
  Fifty years later,  when the Standard Model  (SM) seems to be unbeatable in explaining particle physics data, looking back at that bold hypothesis and big leap in knowledge provides interesting lessons for the future. 

The first point to note is that, despite its many successes, the SM is only a low-energy effective theory. 
In fact, this statement is likely to apply to any four-dimensional quantum field theory (QFT) at arbitrarily high energies, or infinitesimally small distances, given the problem of quantizing gravity in the ultraviolet. However, there are various specific open problems in the SM that call for the existence of new degrees of freedom at high energies, or the embedding of this model into an extended  QFT, with more fields and new symmetries. Two notable examples are the instability of the Higgs sector and the origin of the observed flavor hierarchies. 

Probably the most clear indication that the SM in its original formulation, as classically renormalizable QFT,
should be viewed as an effective QFT is provided by neutrino masses. The experimental evidence of neutrino masses is completely analog to the evidence of CP violation in the SM with two families, which led KM to postulate the existence of a third generation: a clear clue of unknown high-energy dynamics that manifests itself, at low energies, via a new interaction described by a local higher-dimensional operator. More generally,  looking at the (in)consistencies of the SM viewed as an effective QFT, the so-called SMEFT (see e.g.~\cite{Brivio:2017vri,Isidori:2023pyp})  provides interesting clues about its possible ultraviolet completion. In this paper I will focus of the general conclusions that can be derived from flavor physics that, as already fifty years ago, is a precious source of information in this respect. 
I will then focus on an interesting class of extensions of the SM, where the generic clues about the origin of flavor inferred from the effective theory approach find an explicit realisation in terms of new symmetries and new fields. 

\section{Accidental flavor symmetries and modern KM lessons}

A key concept in any effective QFT is that of {\em accidental symmetries},
i.e. symmetries that arise in the lowest-dimensional operators (or the renormalizable part of the QFT)
as indirect consequences of the field content and the symmetries explicitly imposed on the theory
(such as the gauge symmetries).
Within the SMEFT, two well-known examples are the total baryon number, $B$, and total lepton number, $L$.
These are exact accidental global symmetries of the $d=4$ part of the Lagrangian:
they do not need to be imposed in the SM because gauge invariance forbids to write any 
$d=4$~operator violating $B$~or~$L$. 

If the accidental symmetries are not respected by the ultraviolet (UV) completion of the theory,
we expect them to be violated by the higher-dimensional operators. In this context we can 
address the problem of neutrino masses within the SM. Given filed content and gauge symmetries, 
neutrino mass terms are forbidden at $d=4$; however, a neutrino Majorana mass term 
(involving only left-handed neutrinos) can be included via the following dimension-five $L$-violating 
operator~\cite{Weinberg:1979sa}:\footnote{For simplicity, $\SU(2)_L$ and flavor indices are omitted,
 and $v\approx 246$~GeV denotes the Higgs vacuum expectation value.}
\begin{align}
\frac{1}{\Lambda_L}    \bar{\ell_L}^{c} \ell_L H H \,  
 \quad \longrightarrow \quad m_\nu \sim  \frac{  v^2  }{ \Lambda_L}\,.
\label{eq:Weinberg-Operator}
\end{align}
The smallness of neutrino masses can then be attributed to a high value of the effective 
scale $\Lambda_{L}$ characterising $L$-violating interactions. The high value of  $\Lambda_{L}$
(and the even higher scale characterising $B$-violating interactions, 
which are strongly suppressed by the experimental bounds on proton decay) 
is not in contradiction with the possibility of having a lower cutoff scale 
for the SMEFT operators preserving $B$~and~$L$ (starting at  $d=6$), since the symmetry-preserving sector cannot 
induce violations of the global symmetries. In other words, accidental global symmetries allow us to 
define a stable partition of the tower of effective operators into different sectors characterised 
by different cutoff scales, reflecting a possible multi-scale structure of the underlying theory.
The key point is that this partition is stable with respect to quantum corrections. 

Besides $B$~and~$L$, the SM Lagrangian has only two additional exact accidental
global symmetries, corresponding to the conservation of each individual lepton flavor. 
However, a much larger number of {\em approximate accidental symmetries}
appears in the limit where we neglect the tiny Yukawa couplings of the light families: these 
approximate accidental  symmetries are at the origin of what are usually referred to as 
the {\em flavor puzzles}. 

\subsection{The SM flavor puzzle}
The first puzzle, sometimes denoted as the SM flavor puzzle, is why these approximate symmetries are there in
first place. To better quantify the problem, let's give a closer look to the Yukawa interaction of the SM,
\be
-\cL_{\rm Yukawa}=
(Y_u)_{ab}~{\bar q_L^a} H_c u_R^b +
(Y_d)_{ab}~{\bar q_L^a} H d_R^b +
(Y_e)_{ab}~{\bar \ell_L^a} H e_R^b +{\rm h.c.}
\label{eq:Yuakwa}
\ee
The three Yukawa couplings ($Y_{u,d,e}$) are $3\times 3$ complex matrices ($a,b=1 \ldots3$).
In absence of  symmetry principles or dynamical explanations, one would expect their entries to be all
$O(1)$ numbers, while experiments show they are highly non generic.
Via the singular value decomposition, the Yukawa couplings can be put in the form  $Y_{f} = U^\dagger_f \lambda_f V_f$, where $U_f$ and $V_f$ are unitary, while $\lambda_f$ are diagonal.
The nine eigenvalues in $\lambda_{u,d,e}$ exhibit a highly hierarchical 
structure, reflecting the smallness of all fermion masses beside those of the third generation.
In the limit where we neglect entries smaller than $10^{-2}$, they assume the form  
\be 
\lambda_u \approx {\rm diag} (0,0, y_t)\,, \qquad  
\lambda_d \approx {\rm diag} (0,0, y_b)\,, \qquad  
\lambda_e \approx {\rm diag} (0,0, y_\tau)\,, 
\label{eq:lambdas}
\ee
with
\be
y_t =  \frac{\sqrt{2} m_t}{ v} \approx 0.96\,, \qquad 
y_b = \frac{\sqrt{2} m_b}{ v} \approx 0.02\,,  \qquad 
y_\tau  = \frac{\sqrt{2} m_\tau}{ v} \approx 0.01\,.
\ee
Among the unitary matrices, the only physical combination is the 
Cabibbo-Kobayashi-Maskawa (CKM) matrix~\cite{Cabibbo:1963yz,Kobayashi:1973fv}
\be
V_{\rm CKM} = U_u^\dagger U_d \equiv 
\left(\ba{cc|c} V_{ud} & V_{us} & V_{ub} \\  V_{cd} & V_{cs} &  V_{cb} \\ \hline  V_{td} & V_{ts} & V_{td} \ea \right)\,,
\ee
which also exhibits a hierarchical structure with small off-diagonal entries. 
Actually in the the limit where we assume the approximate form in Eq.~(\ref{eq:lambdas}) for the Yukawa 
eigenvalues, the $2\times 2$ light-families block of $V_{\rm CKM}$ becomes unphysical: it can always be rotated 
to the $2\times 2$ identity matrix. In this limit, the only  non-vanishing off-diagonal entry of  $V_{\rm CKM} $ is a single 
light-heavy mixing term, of size 
\be
\epsilon \equiv \sqrt{ |V_{ts}|^2 +  |V_{td}|^2 } \approx |V_{ts} |  \approx 0.04.
\label{eq:epsU2}
\ee

The tiny value of $\epsilon$  is the only source of breaking of the 
\be
\UU(2)^5 = \UU(2)_q \times \UU(2)_u \times \UU(2)_d \times \UU(2)_\ell \times \UU(2)_e
\ee
accidental flavor symmetry implied by Eq.~(\ref{eq:lambdas}), i.e.~the (approximate) 
flavor symmetry resulting from the smallness of light-generation masses~\cite{Barbieri:2011ci,Isidori:2012ts}.
 As already emphasised, this symmetry has no explanation in the SM, or better the SMEFT: it signals a non-trivial flavor structure behind this effective theory.
As discussed in the next section, a natural explanation is that the global $\UU(2)^5$ symmetry arises accidentally in a UV completion of the SM  with extended gauge symmetries, where the charges of the light families forbid the Yukawa interaction at the $d=4$ level.

The smallness of $\epsilon$ allows us to highlight a first important lesson from the KM paper. The $2\times2$ light-family block of the CKM matrix is well separated from the rest. The size of unitarity violations due to light-heavy mixing is $\epsilon^2/2 < 10^{-3}$ (actually even smaller if one looks only at the first raw). It was impossible to detect this effect at KM time
(this is one of the reasons why their hypothesis was highly speculative), but it is not possible even with the precision measurements performed today! The only way to deduce the presence of a third generation, via low-energy experiments involving only light quarks, is to look at specific symmetry-violating effects, such as CP violation in the kaon system, that is exactly what KM did. The general lesson to learn from this observation is that one should not be discouraged by the consistency of many precision SM tests (such as the CKM unitarity test): new degrees of freedom (and in particular new heavy fermions) could still be around the corner. The most efficient way to search for them is to look for violations of the exact and approximate accidental symmetries of the $d=4$ sector of the SMEFT. 

\subsection{The New Physics flavor puzzle}

The second puzzle, sometimes denoted as the New Physics (NP) flavor puzzle, is why in the SMEFT we do not observe 
any significant deviation of the approximate flavor symmetries present in the $d=4$ sector of the theory.
The approximate  $\UU(2)^5$ flavor symmetry is responsible for the smallness of flavor-changing neutral-current (FCNC) processes, such as $B$--$\overline{B}$ and $K$--$\overline{K}$ mixing, which are severely constrained by data. 
Despite the precision and the energy scales involved are very different, the situation is 
similar to that of $B$~and~$L$: the experimental bounds on FCNC processes 
imply high cutoff scales for the $d=6$~operators violating the approximate SM flavor symmetries. 

\begin{figure}[t]
\centering
\includegraphics[width=0.5\textwidth]{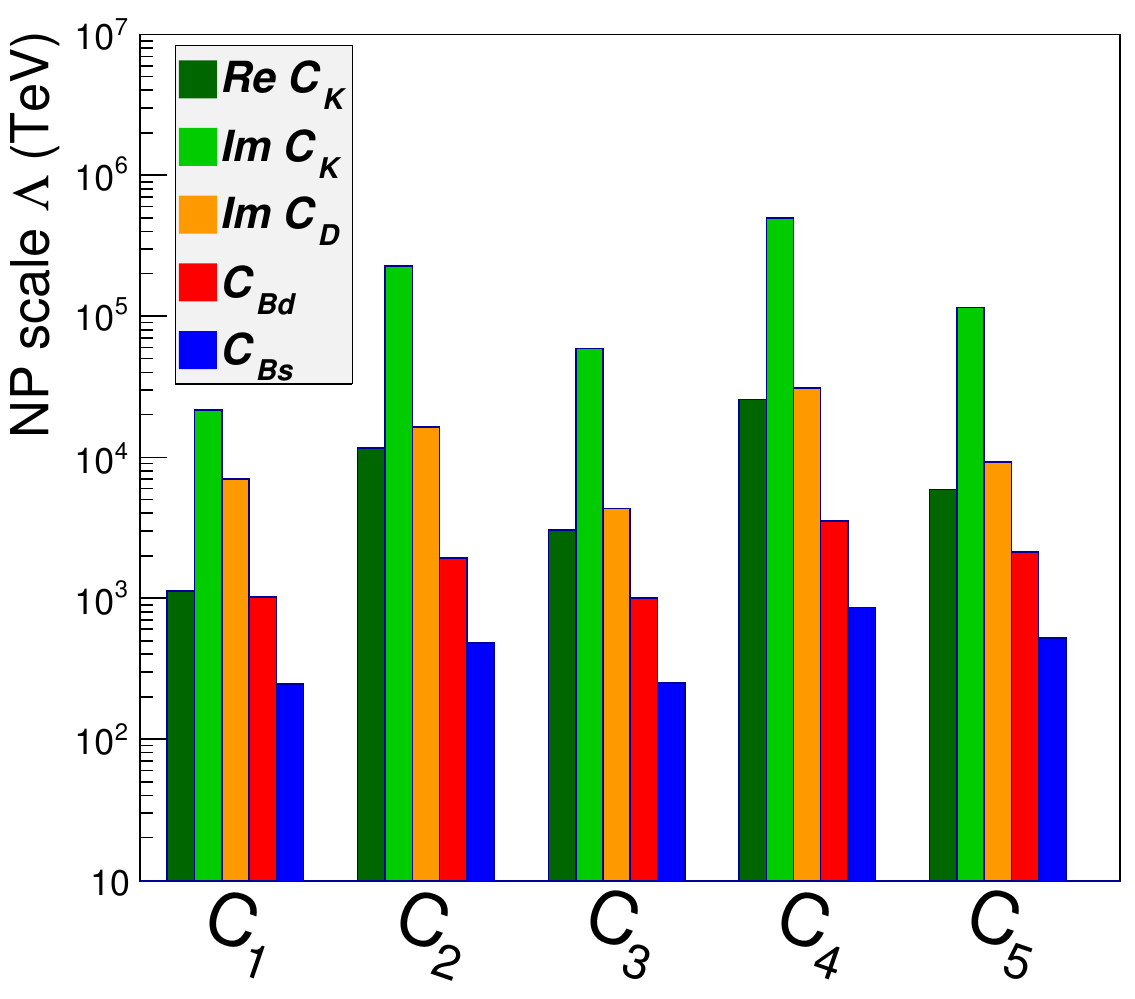}
\vspace{-0.3cm}
\caption{Bounds on the scale of different four-fermion operators contributing to 
$|\Delta F|=2$ amplitudes at the tree level~\cite{Alpigiani:2017lpj}. 
For all the operators an effective coupling $1/\Lambda^2$ is assumed:
$C_1$ denotes the coupling of the left-handed operator in Eq.~(\ref{eq:dfops}), while the definition of the 
other operators can be found in Ref.~\cite{UTfit:2007eik}. 
The different colors denote the different type of $|\Delta F|=2$  transitions used to set the bounds. 
  \label{fig:DF22bounds}}
\end{figure}

To better quantify the problem, let's consider the following part of the SMEFT Lagrangian at $d=6$
containing four-quark operators that can contribute at the tree-level to meson-antimeson mixing
processes (or~$\Delta F=2$ amplitudes):
\be
\Delta \cL^{\rm NP}_{|\Delta F|=2} = \sum_{\{ab\}}  \frac{c_{ab} }{\Lambda^2} Q^{\rm LL}_{ab}~, \qquad 
Q^{\rm LL}_{ab} = (\bar q_L^a \gamma^\mu q_L^b )^2~.
\label{eq:dfops}
\ee
Here $a,b$ are flavor indexes in the basis where $Y_d$ is 
diagonal. In such basis, a given $Q^{\rm LL}_{ab}$ contributes at the tree-level to 
a specific down-type meson-antimeson mixing process: $Q^{\rm LL}_{31}$ 
contributes at the tree-level to $B_d$ mixing,  $Q^{\rm LL}_{32}$  to 
$B_s$ mixing, and so on. In the following we keep the discussion 
generic treating at the same time all down-type meson-antimeson 
mixing amplitudes. All these amplitudes are measured with good accuracy 
and non-SM contributions, if present, are necessary subleading with respect to the SM ones.
For left-handed down-type $\Delta F=2$ amplitudes, the leading SM amplitude is also easy to evaluate 
and can be generically written as 
\be
\cM_{|\Delta F|=2}^{\rm SM} ~ \approx ~ \frac{G_F^2 m_t^2}{16 \pi^2} 
 (V_{ta}^*V_{tb})^2 \langle M_{ab} |   Q^{\rm LL}_{ab}  | {\overline M}_{ab} \rangle \times F(x_t)
\label{eq:gaugeless}
\ee
where $F(x_t= m_t^2/ m_W^2)$ is an order one loop function~\cite{Inami:1980fz}.
From the simple requirement   $| \cM_{|\Delta F|=2}^{\rm NP}| <  | \cM_{|\Delta F|=2}^{\rm SM} |$
one deduces the following bounds~\cite{Artuso:2022ijh}
\bea
\Lambda < \frac{ 3.4~{\rm TeV} }{| V_{ta}^* V^{\phantom{*}}_{tb}|/|c_{ab}|^{1/2}  }
<  \left\{ \ba{l}  
9\times 10^3~{\rm TeV} \times |c_{21}|^{1/2} \qquad {\rm from} \quad 
K^0-\bar K^0
 \\
4\times 10^2~{\rm TeV} \times |c_{31}|^{1/2} \qquad {\rm from}  \quad
B_d-\bar B_d
 \\
7\times 10^1~{\rm TeV} \times |c_{32}|^{1/2} \qquad {\rm from}  \quad
B_s-\bar B_s \ea
\right. 
\label{eq:boundsDF2}
\eea 
The  scaling of the bounds is dictated by the CKM factor $|V_{ta}^* V^{\phantom{*}}_{tb}|$ 
that controls the leading SM contribution. The origin of this factor can simply understood 
as being the leading $\UU(2)^5$ breaking term present in the SM:
\be
|V_{td}^* V^{\phantom{*}}_{ts}| \approx  \epsilon^2 |V_{us}|\,, \qquad 
|V_{tb}^* V^{\phantom{*}}_{td}| \approx  \epsilon |V_{us}|\,, \qquad 
|V_{tb}^* V^{\phantom{*}}_{ts}| \approx  \epsilon \,.
\ee
A more refined analysis, 
with complete statistical treatment and separate bounds 
for the real and the imaginary parts of the various amplitudes,  considering 
also operators  with different Dirac structure, and including also the $D$--$\bar D$ system,
can be found in Ref.~\cite{Alpigiani:2017lpj}.
Some of the results thus obtained are illustrated in Figure~\ref{fig:DF22bounds}.

A na\"ive interpretation of these bounds is that the scale of physics beyond the SM is very high, beyond the realistic reach of direct searches (as least in the short term). However, a more correct interpretation of these bounds is that the amount of $\UU(2)^5$ breaking is small, also beyond the SM. This breaking can be small because it occurs at high energy scales, but it could also be small since it involves small couplings, or a combination of both effects can be at work. Similarly to the small breaking of $L$, the small breaking of $\UU(2)^5$ allows to perform a (quasi) stable partition of the SMEFT (characterised by number and type of  light fermions), reflecting a possible multi-scale structure of the underlying theory.

A second important general lesson from the KM paper emerges: at that time, a possible explanation for the observation of CP violation in the kaon system was the so-called super-weak model proposed by Wolfenstein~\cite{Wolfenstein:1964ks}. In modern language, the super-weak model is nothing but a $\Delta S=2$ 
four-fermion contact interaction characterised by a very small coupling, or a very large effective scale 
\be
\Lambda_{\rm SW} \sim 10^4~{\rm TeV}.
\ee
Indeed $1/\Lambda^2_{\rm SW}$ is the effective coupling appearing in Eq.~(\ref{eq:gaugeless}) in the kaon case (particularly in the imaginary part of the $K$--$\bar K$ mixing amplitude). We now know that the heavy scale $\Lambda_{\rm SW}$ is only a {\em mirage}.
 As proposed first by KM, the consistent extension of the SM with two families necessary to describe CP violation in the kaon system is the addition of a third chiral family, and this is characterised by the scale $m_t \ll \Lambda_{\rm SW}$. The high value of  $\Lambda_{\rm SW}$ is the result of the small breaking of $\UU(2)^5$ in  Eq.~(\ref{eq:gaugeless}). In a similar fashion, all the apparent heavy scales appearing in Fig.~\ref{fig:DF22bounds} could well be mirages. Bounds on NP scales inferred in a  pure effective-theory framework should  be interpreted as upper bounds below which new degrees appears, rather than physical scales characterising new dynamics.

\section{Flavor non-universal gauge interactions}

Explaining the flavor structure of the SM in terms of some underlying dynamics requires extra degrees 
of freedom in the UV that couple to the Higgs field, which is responsible for the Yukawa interaction. 
These new massive degrees of freedom unavoidably contribute to destabilise the electroweak scale, given the 
quadratic sensitivity of the Higgs mass term (the only dimension-2 operator in the SM Lagrangian) to UV dynamics. 
This is the reason why a large fraction of the model-building attempts proposed in the past  
tried to disentangle the (SM) flavor problem and stabilisation of the electroweak scale 
(or the stabilisation of the Higgs mass parameter).

The basic idea of these attempts was  to stabilise the Higgs sector just above the electroweak scale, via some sort of flavor-universal new physics (such as supersymmetry or Higgs-compositeness), while postponing the origin of flavor dynamics to higher scales. The effective-theory construction corresponding to this approach is the so-called Minimal Flavor Violation hypothesis~\cite{DAmbrosio:2002vsn}. 
This scale separation was (and still is) possible because the flavor hierarchies are associated to marginal
operators (the Yukawa interaction), which do not indicate a well-defined energy scale. However, the hypothesis of 
flavor-universal dynamics just above the electroweak scale has become less and less natural in the last few years:
the absence of direct signals of new physics at the LHC has pushed well above 1~TeV the bounds on new degrees of freedom with $O(1)$  flavor-universal couplings to the SM fields. This fact unavoidably worsen the stabilisation of the electroweak scale, independently of the possible solution to the SM flavor problem.

An important point to notice is that the stringent bounds from direct searches are not derived from direct couplings of the new physics to the Higgs or the top quark, but rather by its couplings to the light SM fermions, which play a minor role in the stability of the Higgs mass. As illustrated in the previous section, a similar conclusion applies to the indirect constraints on new physics couplings derived from precision low-energy measurements: are the effective operators involving light SM fermions to be strongly constrained, not those involving only third-generation fields. These general observations seems to indicate that 
flavor and electroweak problems are interrelated and should not be addressed separately, but rather in combination.

\begin{figure}[t]
\centering
\includegraphics[scale=0.45]{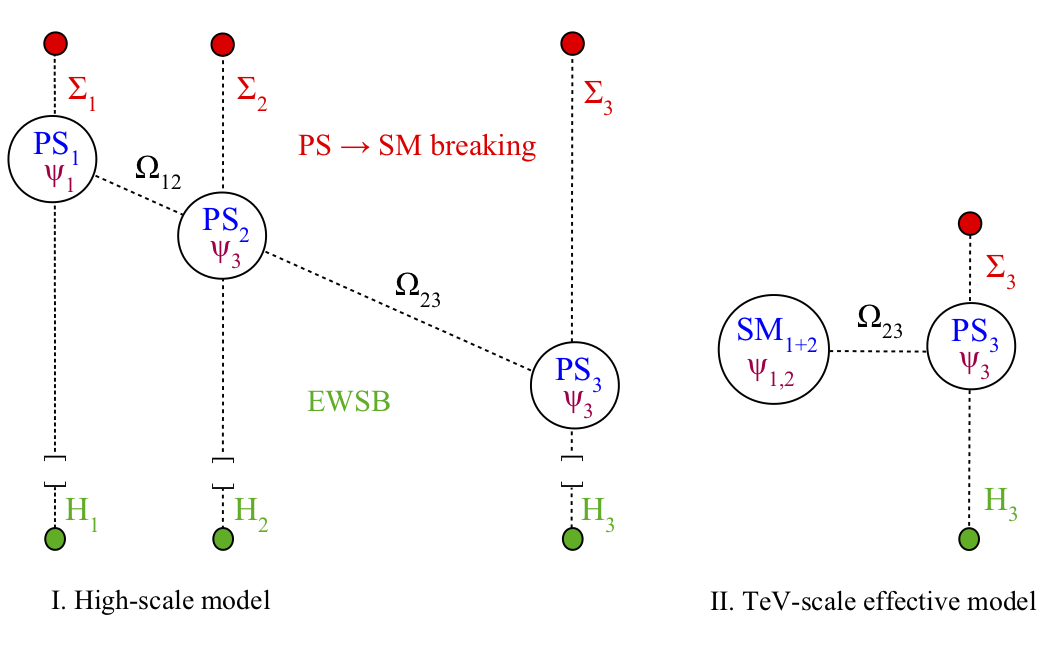}
\caption{Schematic representation of the spontaneous symmetry breaking structure 
${\rm PS}^3 \to$~SM~\cite{Bordone:2017bld,Fuentes-Martin:2020pww}: 
each dotted line denotes a field (or set of fields) with non-vanishing vacuum expectation values. 
Long (short) lines qualitatively indicate small (large) vacuum expectation values, respectively.}
\label{fig:PS3breaking}
\end{figure}

This reasoning is at the base of the growing interest on UV completions of the SM with flavor non-universal gauge interactions. The basic idea is addressing the origin of the flavor hierarchies via gauge interactions which are manifestly 
non universal in flavor, or better via the {\em flavor deconstruction} in the UV~\cite{Arkani-Hamed:2001nha,Craig:2011yk}
of the apparent flavor-universal gauge interactions present in the SM. An interesting prototype for this class of UV completions is the ${\rm PS}^3$ model
 proposed in~\cite{Bordone:2017bld}, whose symmetry breaking chain down to the SM is schematically illustrated in 
Figure~\ref{fig:PS3breaking}.\footnote{The ${\rm PS}^3$ gauge groups corresponds to an independent copy of the semi-simple Pati-Salam gauge group~\cite{Pati:1974yy}, $\SU(4)\times \SU(2)_L\times \SU(2)_R$,  acting separately  on each fermion family.} 
After this attempt, various closely related 
proposals and a few interesting alternatives 
have been discussed in the recent literature~\cite{Greljo:2018tuh,Fuentes-Martin:2020pww,Fuentes-Martin:2020bnh,Fuentes-Martin:2022xnb,FernandezNavarro:2022gst,FernandezNavarro:2023rhv,Davighi:2022fer,Davighi:2022bqf,Davighi:2023iks,Davighi:2023evx}.  
Most of these constructions have been phenomenologically motivated by  the 
so-called $B$-meson anomalies. However,  their interest goes well 
beyond this phenomenological aspect~\cite{Davighi:2022fer,Davighi:2023iks}: the key common feature is the 
appearance of the global $\UU(2)^5$ flavor symmetry as accidental symmetry of the extended gauge sector.

The basic mechanism works as follows~\cite{Davighi:2023iks}. Consider the following gauge symmetries acting on all the SM 
fermions:
$\SU(2)_L$, $\UU(1)_{B-L}$, and $\UU(1)_{R}\equiv U(1)_{T^3_R}$.
Deconstructing in flavor each of these groups,\footnote{
We do not consider the color group, since it does not act on the leptons. Hypercharge is decomposed into a chiral component, $T^3_R$, and a vector-like component, $B-L$ (under which the Higgs is neutral), via the relation $Y= T^3_R +(B-L)/2$. 
The notation $G^{[i]}$ indicate group $G$ acting on the generation $i$.} the Yukawa couplings allowed at the 
renormalisable level ($d=4$) have the following pattern:

\vskip - 0.2 cm
\begin{table}[h]
\begin{center}
\begin{tabular}{cccc}
& 	{\footnotesize $\SU(2)_L^\light \times \SU(2)_L^\heavy$}  
& 	{\footnotesize $\UU(1)_{B-L}^\light \times \UU(1)_{B-L}^\heavy$}
&	{\footnotesize $\UU(1)_R^\light \times \UU(1)_R^\heavy$} \\
&	{\footnotesize [with $H \sim {\bf (1, 2)}$]}
& &	{\footnotesize [with $H\sim (0,-1/2)$]}  \\[2mm]
$Y_f \sim$ 
&	$\qquad\begin{pmatrix} 0&0&0\\ 0&0&0\\ \times&\times&\times \end{pmatrix}\qquad$
&	$\qquad\begin{pmatrix} \times&\times&0\\ \times&\times&0\\ 0&0&\times \end{pmatrix}\qquad$
&	$\qquad\begin{pmatrix} 0&0&\times\\ 0&0&\times\\ 0&0&\times \end{pmatrix}\qquad$ 
\end{tabular}
\end{center}
\end{table}

\vskip - 0.7 cm
\noindent
As can be seen, deconstructing any pair of these groups implies that only the $(Y_f)_{33}$ entries of the Yukawa couplings 
are allowed, hence  a global 
$U(2)^5$ emerges as accidental symmetry of the gauge sector. 
The various options in the different models presented in the literature differ 
essentially on which of these groups are chosen, on the final embedding in the UV, and on
the symmetry breaking pattern. Choosing $\UU(1)_{B-L}$ and $\UU(1)_{R}\equiv U(1)_{T^3_R}$, 
merged into a family-dependendent hypercharge, corresponds to the 
minimal option~\cite{FernandezNavarro:2023rhv,Davighi:2023evx}.

\begin{figure}[t]
\centering
\includegraphics[width=0.8\textwidth]{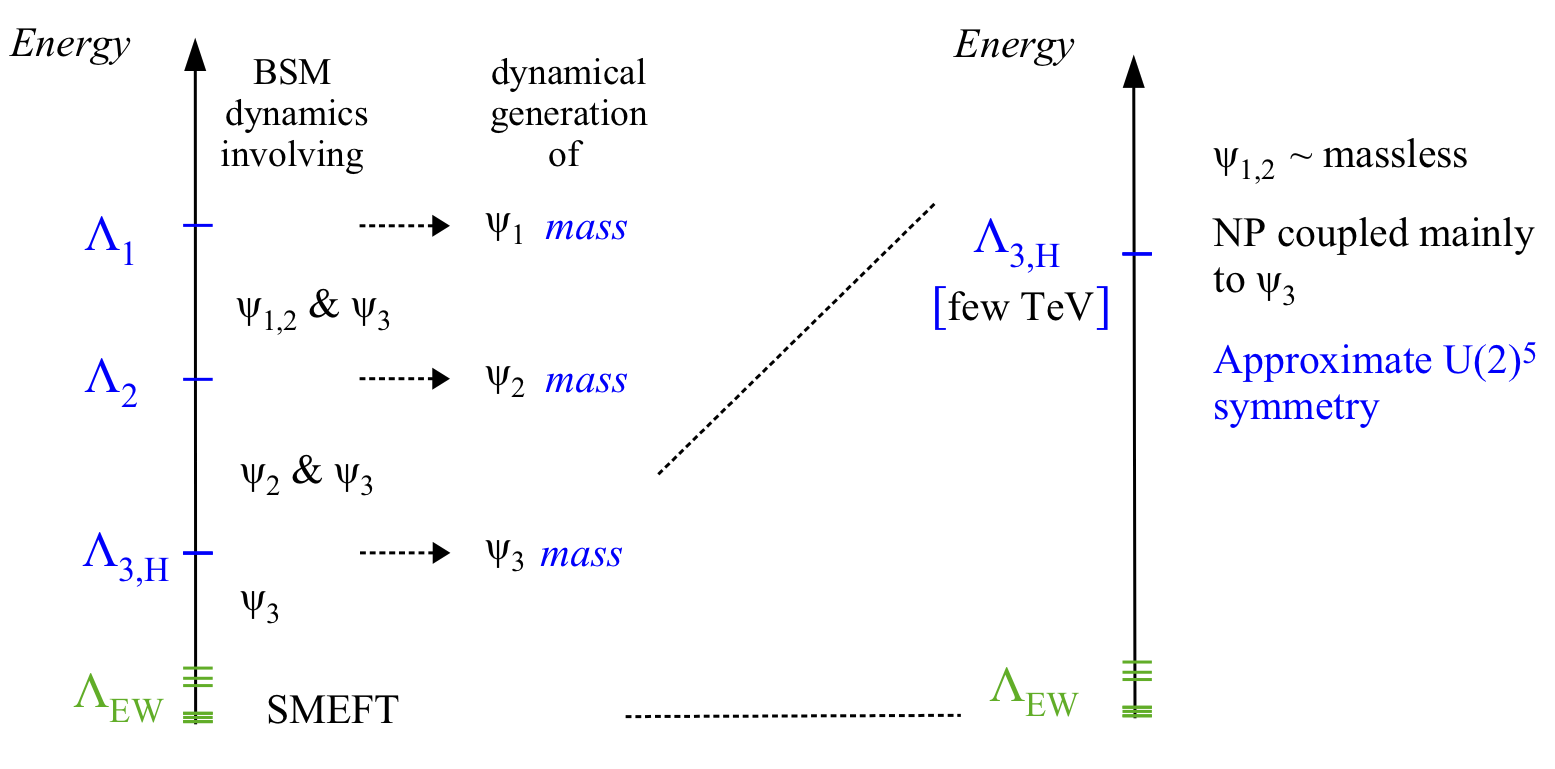} 
\caption{\label{fig:Multiscale} Schematic representation of the possible multi-scale construction at the origin of the SM flavor hierarchies.  }
\end{figure}

This approach also connects to the older, and more general, idea of a UV completion of the  SM based on a (flavored) multi-scale construction~\cite{Dvali:2000ha,Panico:2016ull,Allwicher:2020esa,Barbieri:2021wrc}, 
where the masses for the light generations are generated at increasingly higher scales
 (see Figure~\ref{fig:Multiscale}).
In this context the naturalness of the Higgs mass can be preserved, or, more accurately, the tuning is minimized, if the first layer of new physics enters near the TeV scale, with subsequent layers separated by a few orders of magnitude~\cite{Allwicher:2020esa}. This of course does not rule out the possibility that supersymmetry or compositeness 
play a role in screening the scalar sector from dynamics at even heavier scales; however, if utilised, these general stabilisation mechanisms  could manifest at higher scales given the low-scale stabilisation provided by the lighter flavor non-universal layer of new-physics.

\section{Present anomalies and future prospects}

As shown in~\cite{Davighi:2023iks}, if the flavor-deconstruction idea presented above is implemented in models with a 
semi-simple embedding in the UV, addressing also the hypercharge quantisation problem, the number of possible options is 
quite limited. Most important, all viable models share common features.  A notable common prediction is the expectation 
of a TeV-scale vector leptoquark ($U_1$ field), coupled mainly to the third generation, resulting from a 
unification \`a la Pati and Salam for the third family.

A $U_1$ field coupled mainly to the third generation can explain many of the  anomalies 
observed recently in semileptonic $B$-meson decays~\cite{Alonso:2015sja,Calibbi:2015kma,Barbieri:2015yvd,Bhattacharya:2016mcc,Buttazzo:2017ixm}.
These seemingly coherent set of deviations from the SM predictions, collectively denoted as $B$-meson anomalies, accumulated over the last ten years. The level of significance of one 
specific sub-set, namely the violations of $e/\mu$ universality in $b\to s\ell\bar\ell$ transitions, has suddenly decreased in 2022 after the latest analysis of this effect by LHCb~\cite{LHCb:2022zom}. However, the overall picture remains quite interesting.
As I will briefly illustrated below, they could well be the  first concrete manifestation of the theoretical ideas illustrated in the previous section. 
 
\subsection{$B$-physics anomalies: present status and role of the $U_1$ leptoquark}

The anomalies so far observed can be grouped into three categories, associated to three different underlying amplitudes:

\begin{enumerate}
\item[I.]{\em Violations of $\tau/\mu(e)$ universality in $b\to c\ell\bar\nu$ decays.}\\
This is the oldest anomaly (it was observed first by Babar in 2012~\cite{BaBar:2012obs}) and the one where more experiments are involved. The main observables are the Lepton Flavor Universality (LFU) ratios $R_D$ and $R_{D^*}$, whose SM predictions 
have small (and well controlled) theoretical errors. 
According to HFLAG~\cite{HeavyFlavorAveragingGroup:2022wzx}, the latest global fit indicate an enhancement 
over the SM predictions of the tau rates (over light leptons) of about 10-15\%, with a significance of 3.2$\sigma$ (see Figure~\ref{fig:RD}).
The SM contribution to these processes appears already at the tree level, and is suppressed only by $|V_{cb}|$. 
Hence a NP interpretation of this anomaly necessary points toward a sizeable amplitude, corresponding to an
effective scale of at most a few TeV.

\begin{figure}[t]
\centering
\includegraphics[scale=0.65]{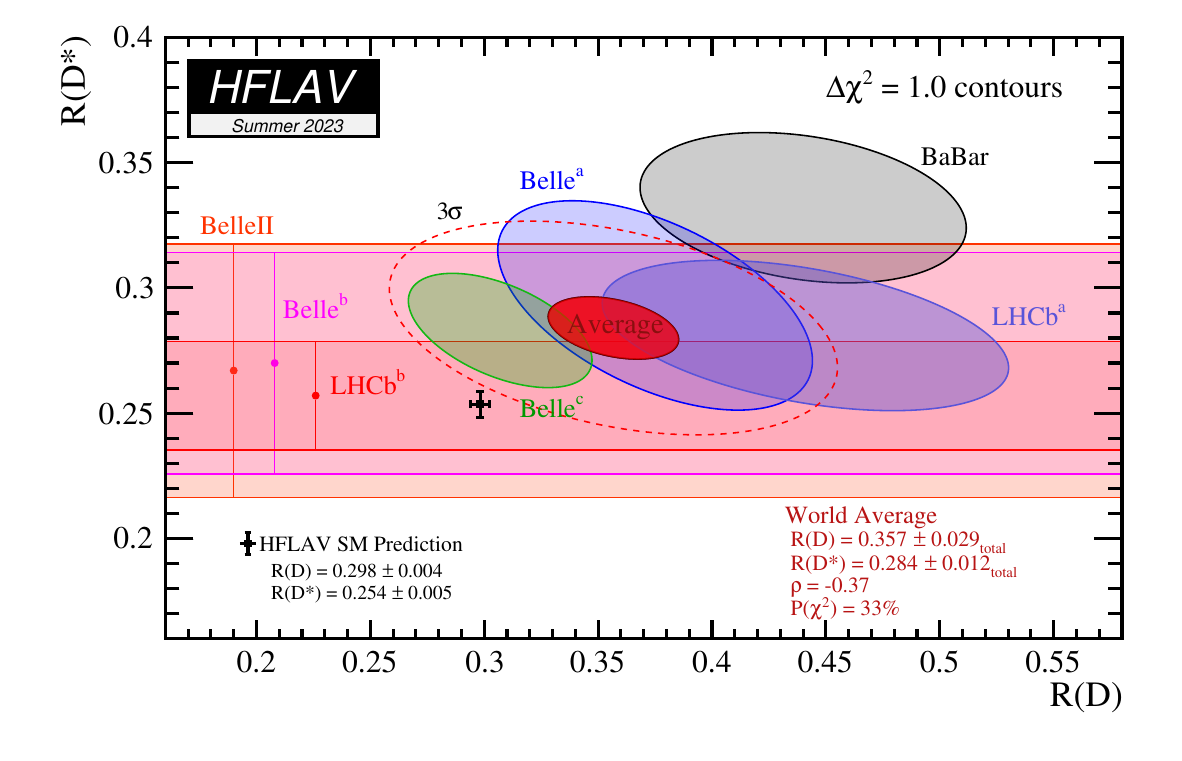}
\caption{Status of the experimental results on the LFU ratios $R_D$ and $R_{D^*}$  (colored bands and ellipses) vs.~SM predictions (black cross) in summer 2023~\cite{HeavyFlavorAveragingGroup:2022wzx}. The dotted red ellipse indicate the $3\sigma$ countour resulting from the global average of the experimental results. }
\label{fig:RD}
\end{figure}

\item[II.]{\em Lepton-universal anomaly in $b\to s\ell\bar\ell$ decays.}\\
Rates and differential distributions of several exclusive and inclusive $b\to s\ell\bar\ell$ transitions  (most notably $B\to K \ell\bar\ell$, $B\to K^{*} \ell\bar\ell$, and $B_s\to \phi \ell\bar\ell$ decays) are in significant tensions with the corresponding SM predictions
(see e.g.~Ref.~\cite{Alguero:2023jeh, Gubernari:2022hxn, Altmannshofer:2021qrr,Hurth:2020ehu} for recent analyses). The effect is well described by a short-distance $b\to s\ell\bar\ell$ amplitude, with the two leptons 
in a vector state.\footnote{In the standard notation for the $b\to s\ell\bar\ell$ 
effective Lagrangian, this is described as a $\approx 20\%$ suppression 
of the effective coefficient $C_9$ compared to its SM value~\cite{Alguero:2023jeh,Gubernari:2022hxn, Altmannshofer:2021qrr,Hurth:2020ehu}.}
This anomaly is the most robust  from a pure experimental point of view; 
however,  doubts about the reliability of the theory errors have been raised in~\cite{Ciuchini:2019usw}, given
the difficulty of estimating charm rescattering  effects in exclusive hadronic processes. The consistency of the results
obtained at low and high dilepton invariant mass ($q^2$),
and in different hadronic modes, provides an important consistency check of the short-distance origin of the effect
(hence a possible NP interpretation). A further evidence in favor of the short-distance hypothesis 
has been obtained recently by a semi-inclusive analysis at high $q^2$~\cite{Isidori:2023unk}. 
The significance of the anomaly range from the very conservative $2\sigma$ obtained in~\cite{Isidori:2023unk},
using only on semi-inclusive data, to the $5\sigma$ obtained in~\cite{Alguero:2023jeh} from global fits.

\item[III.]{\em LFU tests in $b\to s\ell\bar\ell$ decays and $B_s\to \mu^+\mu^-$.}\\
Till 2022, a statistically significant violation of $\mu/e$ universality in $b\to s\ell\bar\ell$ decays
was hinted by the $15-20\%$ suppression of the
LFU ratios $R_K$ and $R_{K^{*}}$ relative to the clean SM prediction 
$R^{\rm SM}_{K^{(*)}}\approx1$~\cite{Bordone:2016gaq,Isidori:2020acz}.
This effect was consistent with the  suppression of $\cB(B_s\to \mu^+\mu^-)$ compared to its SM value. 
As anticipated, the significance of these anomalies has been drastically reduced by the 
LHCb re-analysis in~\cite{LHCb:2022zom} and, to a lesser extent, also by a new  
CMS analysis of $\cB(B_s\to \mu^+\mu^-)$~\cite{CMS:2022mgd}. Present data are fully compatible with the SM;
however, violations of $\mu/e$ universality in $b\to s\ell\bar\ell$  up to $10\%$ cannot
be excluded.
\end{enumerate}

\begin{figure}[t]
\centering
\includegraphics[scale=0.45]{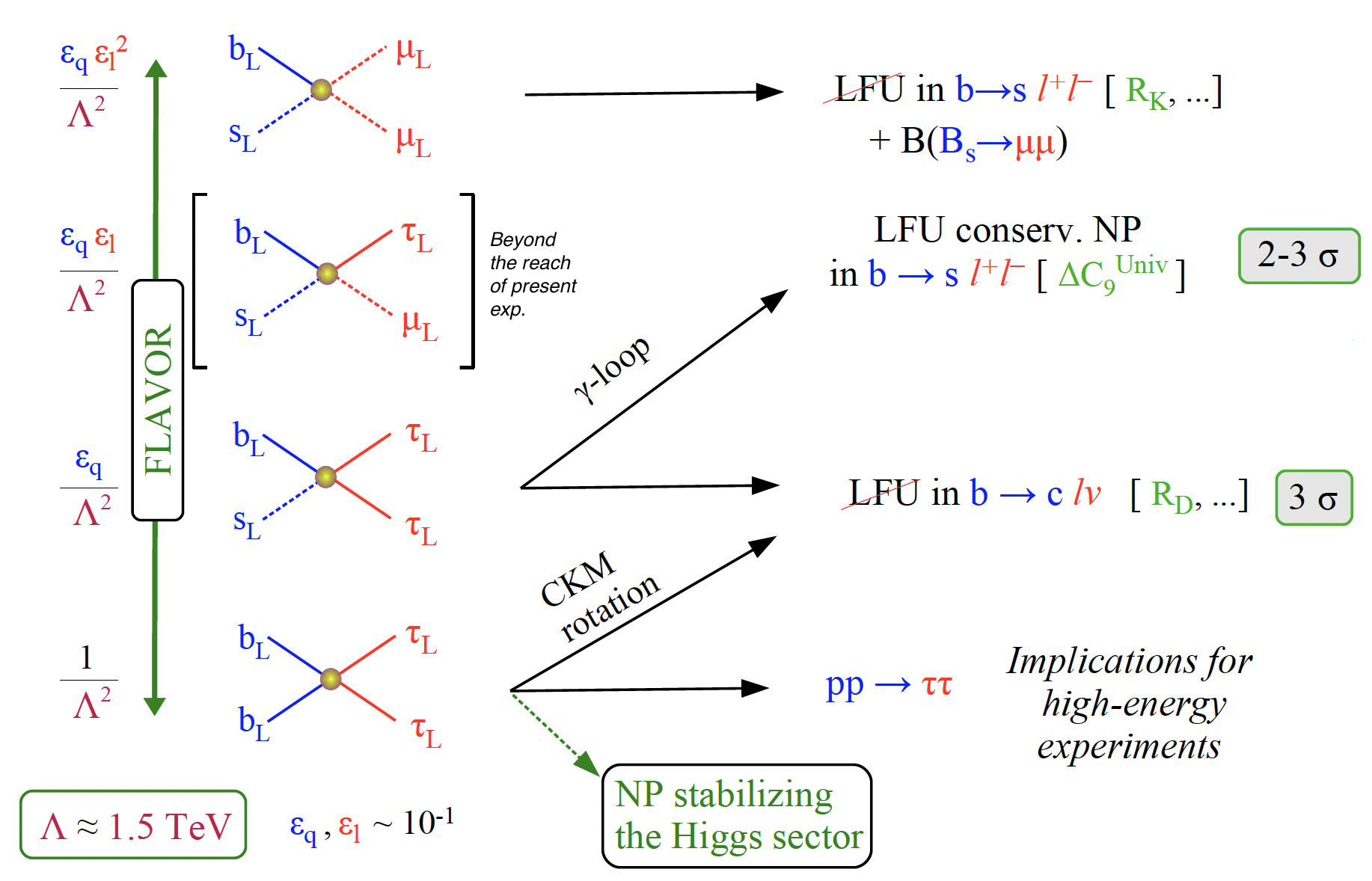}
\caption{Schematic (semi-qualitative) interpretation of the $B$-meson anomalies in terms of semi-leptonic effective operators.
The arrow denoted {\em CKM rotation} denotes the effect obtained after rotating left-handed fields into their mass-eigenstate basis.
The arrow denoted  {\em $\gamma$-loop} indicate the $b\to s (\tau\tau \to \ell\bar\ell)$ lepton-universal amplitude obtained 
at the one-loop level~\cite{Bobeth:2011st,Crivellin:2018yvo}.}
\label{fig:anomalies}
\end{figure}

A na\"ive intepretation of the $B$-meson anomalies in terms of semi-leptonic effective operators is illustrated in 
Figure~\ref{fig:anomalies}. The observations are well consistent with a leading interaction 
 of the type $ (\bar b_L \gamma^\mu b_L )( \bar \tau_L \gamma_\mu \tau_L )$, involving only left-handed 
 third-generations fields,\footnote{An effective interaction of the type 
 [$(\bar b_L \gamma^\mu \ell_L )( \bar \tau_R \gamma_\mu \tau_R )$+h.c.],  of similar strength,
 is also compatible with present data~\cite{Aebischer:2022oqe}.} corresponding to an effective scale 
 $\Lambda \in [1,2]$~TeV~\cite{Buttazzo:2017ixm,Aebischer:2022oqe}.
 Terms with the same electroweak structure where the left-handed third-generations fields are replaced by 
second-generation fields should also be present, but with suppressed 
coefficients. In particular, there is a clear evidence for a suppression factor $\varepsilon_q \sim 10^{-1}$ for each 
second-generation quark filed. A similar suppression factor in the lepton sector (i.e.~$\varepsilon_\ell \sim 10^{-1}$)
implies a $10\%$  violation of  $\mu/e$ universality in $b\to s\ell\bar\ell$ decays. 

The tower of effective operators described above fits well with what expected by the tree-level exchange of 
a $U_1$ leptoquark coupled mainly to the third generation, with $\UU(2)^5$ breaking terms controlled by 
$\varepsilon_{q,\ell}$, as proposed first in~\cite{Barbieri:2015yvd}.
 Note the nice consistency between the value of $\varepsilon_q$  (determined by 
the anomalies I and II), with the value of $\epsilon$ determined in (\ref{eq:epsU2}) 
by the structure of the CKM matrix. 

\subsection{Future prospects}

The coherent hints of a possible unification a la Pati-Salam for the third generation are definitely interesting; however, at present the statistical significance is still quite low. On the other hand, if this hypothesis is correct, more particles and more signals should be accessible in the short term at both low and high energies.

\begin{figure}[t]
    \centering
    \includegraphics[width=0.47\linewidth]{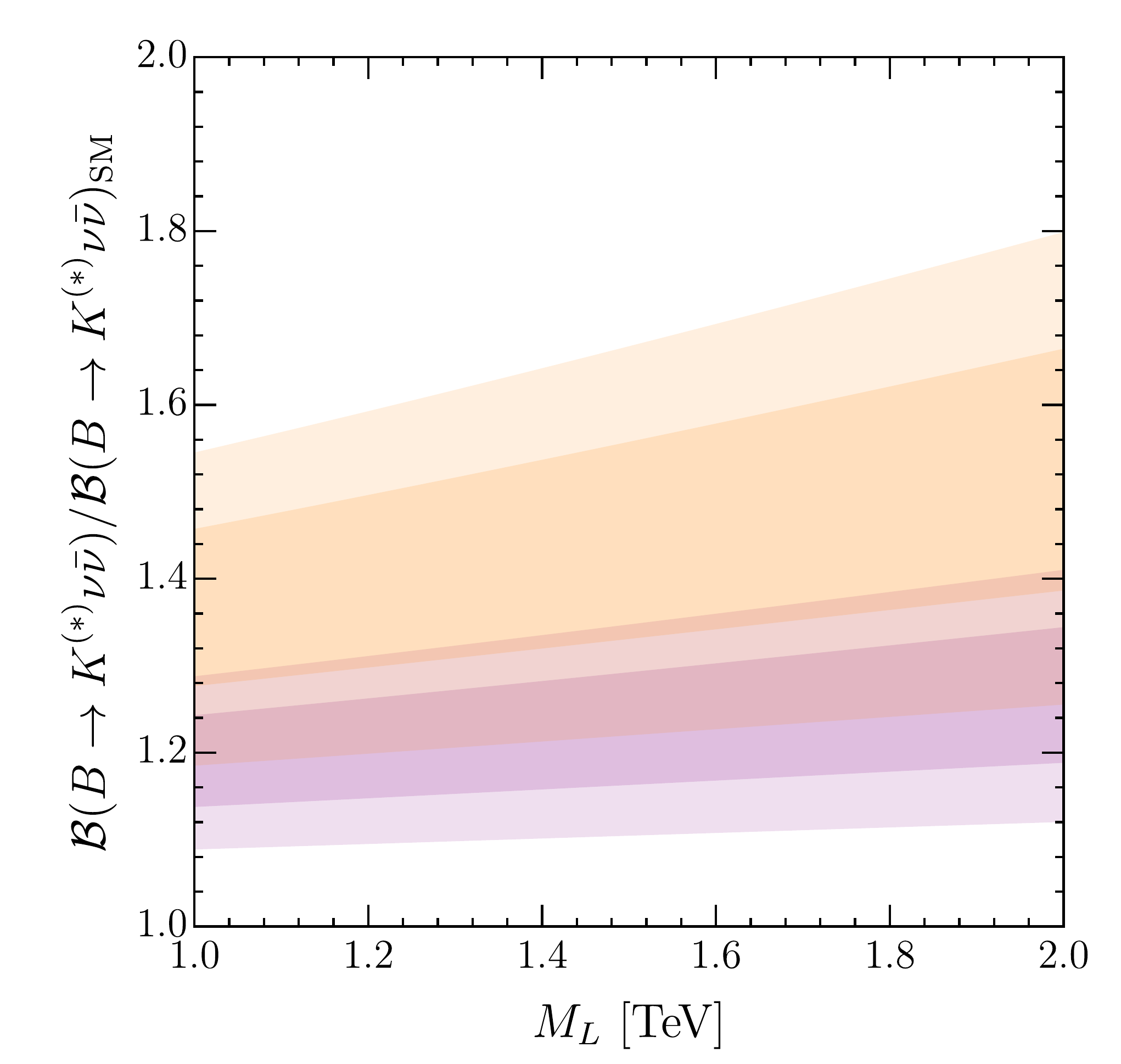} 
    \includegraphics[width=0.45\linewidth]{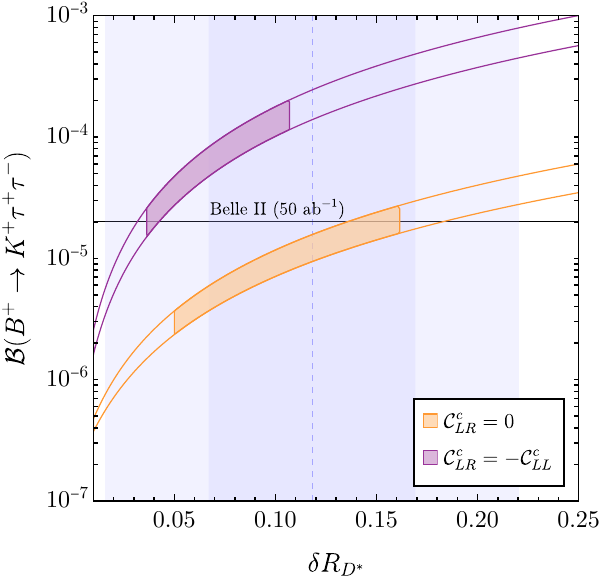}
    \caption{Examples of low-energy observables which exhibit significant deviations from the SM in models with third-family 
    Pati-Salam unification at the TeV scale, addressing the $B$-meson anomalies. Left: $\cB(B\to K^{(*)}\bar \nu \nu)$~vs.~the mass 
    of vector-like leptons~\cite{Fuentes-Martin:2020hvc}.
    Right: $\cB(B\to K^+ \tau^+ \tau^-)$ vs.~the relative shift in $R_{D^*}$ ~\cite{Aebischer:2022oqe}
    (the blue bands denote the present 1 and 2 $\sigma$ value of $R_{D^*}$).
     In both plots the orange (violet) bands correspond to a pure left-handed (left-right symmetric) interaction of the $U_1$
     to third-generation fermions. }
    \label{fig:bstautau} 
\end{figure}

First of all, the $U_1$ leptoquark cannot be alone~\cite{Baker:2019sli}. The minimal consistent gauge group hosting this massive vector 
is $\SU(4)^{[3]}\times \SU(3)^{[12]} \times \SU(2)_L \times \UU(1)$~\cite{DiLuzio:2017vat}. Its breaking down to the SM 
implies the presence of three sets of massive vectors:  in addition to 
the $U_1$,  also a color octect $G$  (denoted {\em coloron}, behaving like a heavy gluon) and a $Z^\prime$ (singlet under the SM) are necessarily present. All these fields are coupled dominantly to third-generation fermions, with similar mass ($M_V$) 
and couplings ($g_4$).
As mentioned above, if we require the $U_1$ to address the $B$-meson anomalies, then 
$M_V/g_4 \in [1,2]$~TeV.  Such range is very close to present bound 
from direct searches. In particular, the most promising channels to detect these massive vectors are  
$pp\to \bar\tau\tau$ ($t$--channel  $U_1$ and $s$--channel $Z^\prime$ exchange~\cite{Faroughy:2016osc,Baker:2019sli}) and $pp\to \bar t t$ ($s$--channel $G$ exchange~\cite{Baker:2019sli,Cornella:2021sby}).

A general expectation of these new states and, more generally, of any model with flavor deconstruction at the TeV scale, 
are deviations from the SM predictions in various electroweak precision observables  (typically at the few per-mil level). 
As shown in~\cite{Davighi:2023evx} by means of an explicit example,  FCC-ee would allow to extensively probe the natural parameter space of such  class of models via electroweak observables. 

Last but not least, a necessary ingredient of this class of models are vector-like fermions (i.e.~fermions where left- and right-handed
components have the same gauge quantum numbers, allowing a Dirac-mass term). These fermions are responsible for the 
heavy-light mixing of the chiral fermions (or the structure of the CKM matrix), as it happens in a wider  class of motivated SM extensions~\cite{Botella:2016ibj}. The interplay of massive vectors and vector-like fermions  give rise to a series of additional
low-energy signatures which are within the reach of the present generation of flavor-phyics experiments (see Figure~\ref{fig:bstautau} ).
 These include lepton-flavor violating decays such as  $B\to K^{(*)}\bar \tau \mu$,  
$\tau \to \phi \mu$,  or $\tau \to 3\mu$ close to present bounds (see e.g.~\cite{Cornella:2021sby});
large enhancements over the SM predictions for rare $B$ decays into $\tau^+\tau^-$ 
 pairs~\cite{Capdevila:2017iqn}; $O(10-30\%)$ deviations over the SM predictions for $\cB(B\to K^{(*)}\bar \nu \nu)$~\cite{Fuentes-Martin:2020hvc}  and  $\cB(K^+ \to \pi^+\bar \nu \nu)$~\cite{Crosas:2022quq}.

\section{Conclusion}

The KM mechanism plays a major role in our understanding of fundamental interactions. 
As I highlighted in this paper, beside the invaluable contribution that KM gave to the construction of the Standard Model,
there are interesting modern lessons we can deduce from their work which are still valuable today, 
when applied to the search for physics beyond the SM.
These range from the importance of accidental symmetries in identifying clues of heavy dynamics 
(beyond a given renormalizable theory) to the importance of building complete UV models.  
A key lesson is also not to be afraid of putting forward bold hypotheses, if they are well motivated and can solve open problems. 
Last but not least, the KM paper is a perfect illustration of the importance of flavor mixing as a window toward 
new dynamics. As it did fifty years ago, flavor physics still hides interesting puzzles and may represent the key to understanding the nature of physics above the electroweak scale.

\newpage

\bibliographystyle{ptephy}
\bibliography{KMpaper}

\begin{thebibliography}{10}

\bibitem{Kobayashi:1973fv}
Makoto Kobayashi and Toshihide Maskawa, Prog. Theor. Phys., {\bf 49}, 652--657
  (1973).

\bibitem{Brivio:2017vri}
Ilaria Brivio and Michael Trott, Phys. Rept., {\bf 793}, 1--98 (2019),
  {{arXiv:1706.08945}}.

\bibitem{Isidori:2023pyp}
Gino Isidori, Felix Wilsch, and Daniel Wyler (2023),  {{arXiv:2303.16922}}.

\bibitem{Weinberg:1979sa}
Steven Weinberg, Phys. Rev. Lett., {\bf 43}, 1566--1570 (1979).

\bibitem{Cabibbo:1963yz}
Nicola Cabibbo, Phys. Rev. Lett., {\bf 10}, 531--533 (1963).

\bibitem{Barbieri:2011ci}
Riccardo Barbieri, Gino Isidori, Joel Jones-Perez, Paolo Lodone, and David~M.
  Straub, Eur. Phys. J. C, {\bf 71}, 1725 (2011),  {{arXiv:1105.2296}}.

\bibitem{Isidori:2012ts}
Gino Isidori and David~M. Straub, Eur. Phys. J. C, {\bf 72}, 2103 (2012),
  {{arXiv:1202.0464}}.

\bibitem{Inami:1980fz}
T.~Inami and C.~S. Lim, Prog. Theor. Phys., {\bf 65}, 297, [Erratum:
  Prog.Theor.Phys. 65, 1772 (1981)] (1981).

\bibitem{Artuso:2022ijh}
Marina Artuso, Gino Isidori, and Sheldon Stone,
\newblock {\em {New Physics in b Decays}},
\newblock  (World Scientific, 2022).

\bibitem{Alpigiani:2017lpj}
Cristiano Alpigiani et~al.,
\newblock {Unitarity Triangle Analysis in the Standard Model and Beyond},
\newblock In {\em {5th Large Hadron Collider Physics Conference}} (10 2017),
  {{arXiv:1710.09644}}.

\bibitem{UTfit:2007eik}
M.~Bona et~al., JHEP, {\bf 03}, 049 (2008),  {{arXiv:0707.0636}}.

\bibitem{Wolfenstein:1964ks}
L.~Wolfenstein, Phys. Rev. Lett., {\bf 13}, 562--564 (1964).

\bibitem{DAmbrosio:2002vsn}
G.~D'Ambrosio, G.~F. Giudice, G.~Isidori, and A.~Strumia, Nucl. Phys. B, {\bf
  645}, 155--187 (2002),  {{hep-ph/0207036}}.

\bibitem{Arkani-Hamed:2001nha}
Nima Arkani-Hamed, Andrew~G. Cohen, and Howard Georgi, Phys. Lett. B, {\bf
  513}, 232--240 (2001),  {{hep-ph/0105239}}.

\bibitem{Craig:2011yk}
Nathaniel Craig, Daniel Green, and Andrey Katz, JHEP, {\bf 07}, 045 (2011),
  {{arXiv:1103.3708}}.

\bibitem{Bordone:2017bld}
Marzia Bordone, Claudia Cornella, Javier Fuentes-Martin, and Gino Isidori,
  Phys. Lett. B, {\bf 779}, 317--323 (2018),  {{arXiv:1712.01368}}.

\bibitem{Pati:1974yy}
Jogesh~C. Pati and Abdus Salam, Phys. Rev. D, {\bf 10}, 275--289, [Erratum:
  Phys.Rev.D 11, 703--703 (1975)] (1974).

\bibitem{Greljo:2018tuh}
Admir Greljo and Ben~A. Stefanek, Phys. Lett. B, {\bf 782}, 131--138 (2018),
  {{arXiv:1802.04274}}.

\bibitem{Fuentes-Martin:2020pww}
Javier Fuentes-Martin, Gino Isidori, Julie Pag\`es, and Ben~A. Stefanek, Phys.
  Lett. B, {\bf 820}, 136484 (2021),  {{arXiv:2012.10492}}.

\bibitem{Fuentes-Martin:2020bnh}
Javier Fuentes-Mart\'\i{}n and Peter Stangl, Phys. Lett. B, {\bf 811}, 135953
  (2020),  {{arXiv:2004.11376}}.

\bibitem{Fuentes-Martin:2022xnb}
Javier Fuentes-Martin, Gino Isidori, Javier~M. Lizana, Nudzeim Selimovic, and
  Ben~A. Stefanek, Phys. Lett. B, {\bf 834}, 137382 (2022),
  {{arXiv:2203.01952}}.

\bibitem{FernandezNavarro:2022gst}
Mario Fern\'andez~Navarro and Stephen~F. King, JHEP, {\bf 02}, 188 (2023),
  {{arXiv:2209.00276}}.

\bibitem{FernandezNavarro:2023rhv}
Mario Fern\'andez~Navarro and Stephen~F. King (2023),  {{arXiv:2305.07690}}.

\bibitem{Davighi:2022fer}
Joe Davighi and Joseph Tooby-Smith, JHEP, {\bf 09}, 193 (2022),
  {{arXiv:2201.07245}}.

\bibitem{Davighi:2022bqf}
Joe Davighi, Gino Isidori, and Marko Pesut, JHEP, {\bf 04}, 030 (2023),
  {{arXiv:2212.06163}}.

\bibitem{Davighi:2023iks}
Joe Davighi and Gino Isidori, JHEP, {\bf 07}, 147 (2023),
  {{arXiv:2303.01520}}.

\bibitem{Davighi:2023evx}
Joe Davighi and Ben~A. Stefanek (2023),  {{arXiv:2305.16280}}.

\bibitem{Dvali:2000ha}
G.~R. Dvali and Mikhail~A. Shifman, Phys. Lett. B, {\bf 475}, 295--302 (2000),
  {{hep-ph/0001072}}.

\bibitem{Panico:2016ull}
Giuliano Panico and Alex Pomarol, JHEP, {\bf 07}, 097 (2016),
  {{arXiv:1603.06609}}.

\bibitem{Allwicher:2020esa}
Lukas Allwicher, Gino Isidori, and Anders~Eller Thomsen, JHEP, {\bf 01}, 191
  (2021),  {{arXiv:2011.01946}}.

\bibitem{Barbieri:2021wrc}
Riccardo Barbieri, Acta Phys. Polon. B, {\bf 52}, 789 (2021),
  {{arXiv:2103.15635}}.

\bibitem{Alonso:2015sja}
Rodrigo Alonso, Benjam\'\i{}n Grinstein, and Jorge Martin~Camalich, JHEP, {\bf
  10}, 184 (2015),  {{arXiv:1505.05164}}.

\bibitem{Calibbi:2015kma}
Lorenzo Calibbi, Andreas Crivellin, and Toshihiko Ota, Phys. Rev. Lett., {\bf
  115}, 181801 (2015),  {{arXiv:1506.02661}}.

\bibitem{Barbieri:2015yvd}
Riccardo Barbieri, Gino Isidori, Andrea Pattori, and Fabrizio Senia, Eur. Phys.
  J. C, {\bf 76}, 67 (2016),  {{arXiv:1512.01560}}.

\bibitem{Bhattacharya:2016mcc}
Bhubanjyoti Bhattacharya, Alakabha Datta, Jean-Pascal Gu\'evin, David London,
  and Ryoutaro Watanabe, JHEP, {\bf 01}, 015 (2017),  {{arXiv:1609.09078}}.

\bibitem{Buttazzo:2017ixm}
Dario Buttazzo, Admir Greljo, Gino Isidori, and David Marzocca, JHEP, {\bf 11},
  044 (2017),  {{arXiv:1706.07808}}.

\bibitem{LHCb:2022zom}
R.~Aaij et~al. (2022),  {{arXiv:2212.09153}}.

\bibitem{BaBar:2012obs}
J.~P. Lees et~al., Phys. Rev. Lett., {\bf 109}, 101802 (2012),
  {{arXiv:1205.5442}}.

\bibitem{HeavyFlavorAveragingGroup:2022wzx}
Yasmine~Sara Amhis et~al., Phys. Rev. D, {\bf 107}(5), 052008 (2023),
  {{arXiv:2206.07501, {\em updates:} https://hflav.web.cern.ch}}.

\bibitem{Alguero:2023jeh}
Marcel Alguer\'o, Aritra Biswas, Bernat Capdevila, S\'ebastien Descotes-Genon,
  Joaquim Matias, and Mart\'\i{}n Novoa-Brunet, Eur. Phys. J. C, {\bf 83}(7),
  648 (2023),  {{arXiv:2304.07330}}.

\bibitem{Gubernari:2022hxn}
Nico Gubernari, M\'eril Reboud, Danny van Dyk, and Javier Virto, JHEP, {\bf
  09}, 133 (2022),  {{arXiv:2206.03797}}.

\bibitem{Altmannshofer:2021qrr}
Wolfgang Altmannshofer and Peter Stangl, Eur. Phys. J. C, {\bf 81}(10), 952
  (2021),  {{arXiv:2103.13370}}.

\bibitem{Hurth:2020ehu}
T.~Hurth, F.~Mahmoudi, and S.~Neshatpour, Phys. Rev. D, {\bf 103}, 095020
  (2021),  {{arXiv:2012.12207}}.

\bibitem{Ciuchini:2019usw}
Marco Ciuchini, Ant\'onio~M. Coutinho, Marco Fedele, Enrico Franco, Ayan Paul,
  Luca Silvestrini, and Mauro Valli, Eur. Phys. J. C, {\bf 79}(8), 719 (2019),
  {{arXiv:1903.09632}}.

\bibitem{Isidori:2023unk}
Gino Isidori, Zachary Polonsky, and Arianna Tinari (2023),
  {{arXiv:2305.03076}}.

\bibitem{Bordone:2016gaq}
Marzia Bordone, Gino Isidori, and Andrea Pattori, Eur. Phys. J. C, {\bf 76}(8),
  440 (2016),  {{arXiv:1605.07633}}.

\bibitem{Isidori:2020acz}
Gino Isidori, Saad Nabeebaccus, and Roman Zwicky, JHEP, {\bf 12}, 104 (2020),
  {{arXiv:2009.00929}}.

\bibitem{CMS:2022mgd}
Armen Tumasyan et~al., Phys. Lett. B, {\bf 842}, 137955 (2023),
  {{arXiv:2212.10311}}.

\bibitem{Bobeth:2011st}
Christoph Bobeth and Ulrich Haisch, Acta Phys. Polon. B, {\bf 44}, 127--176
  (2013),  {{arXiv:1109.1826}}.

\bibitem{Crivellin:2018yvo}
Andreas Crivellin, Christoph Greub, Dario M\"uller, and Francesco Saturnino,
  Phys. Rev. Lett., {\bf 122}(1), 011805 (2019),  {{arXiv:1807.02068}}.

\bibitem{Aebischer:2022oqe}
Jason Aebischer, Gino Isidori, Marko Pesut, Ben~A. Stefanek, and Felix Wilsch,
  Eur. Phys. J. C, {\bf 83}(2), 153 (2023),  {{arXiv:2210.13422}}.

\bibitem{Fuentes-Martin:2020hvc}
Javier Fuentes-Mart\'\i{}n, Gino Isidori, Matthias K\"onig, and Nud\v{z}eim
  Selimovi\'c, Phys. Rev. D, {\bf 102}, 115015 (2020),  {{arXiv:2009.11296}}.

\bibitem{Baker:2019sli}
Michael~J. Baker, Javier Fuentes-Mart\'\i{}n, Gino Isidori, and Matthias
  K\"onig, Eur. Phys. J. C, {\bf 79}(4), 334 (2019),  {{arXiv:1901.10480}}.

\bibitem{DiLuzio:2017vat}
Luca Di~Luzio, Admir Greljo, and Marco Nardecchia, Phys. Rev. D, {\bf 96}(11),
  115011 (2017),  {{arXiv:1708.08450}}.

\bibitem{Faroughy:2016osc}
Darius~A. Faroughy, Admir Greljo, and Jernej~F. Kamenik, Phys. Lett. B, {\bf
  764}, 126--134 (2017),  {{arXiv:1609.07138}}.

\bibitem{Cornella:2021sby}
Claudia Cornella, Darius~A. Faroughy, Javier Fuentes-Martin, Gino Isidori, and
  Matthias Neubert, JHEP, {\bf 08}, 050 (2021),  {{arXiv:2103.16558}}.

\bibitem{Botella:2016ibj}
Francisco~J. Botella, G.~C. Branco, Miguel Nebot, M.~N. Rebelo, and J.~I.
  Silva-Marcos, Eur. Phys. J. C, {\bf 77}(6), 408 (2017),
  {{arXiv:1610.03018}}.

\bibitem{Capdevila:2017iqn}
Bernat Capdevila, Andreas Crivellin, S\'ebastien Descotes-Genon, Lars Hofer,
  and Joaquim Matias, Phys. Rev. Lett., {\bf 120}(18), 181802 (2018),
  {{arXiv:1712.01919}}.

\bibitem{Crosas:2022quq}
\`Oscar~L. Crosas, Gino Isidori, Javier~M. Lizana, Nudzeim Selimovic, and
  Ben~A. Stefanek, Phys. Lett. B, {\bf 835}, 137525 (2022),
  {{arXiv:2207.00018}}.

\end{thebibliography}

\end{document}